\documentclass[twocolumn,showpacs,showkeys,preprintnumbers,amsmath,amssymb,nofootinbib]{revtex4}
\usepackage{graphicx}
\usepackage{dcolumn}
\usepackage{bm}
\begin{document}
\title{\bf  Variational Method for  Photon Emission from Quark-Gluon Plasma }
\author{S. V. Suryanarayana\footnote{ In nuclear physics journals and arxiv listings,  my  name used to appear as S.V.S. Sastry. Hereafter (by deleting
Sastry), my name will appear as, S.V. Suryanarayana.}}
\email{suryanarayan7@yahoo.com}
\email{snarayan@barc.gov.in}
\affiliation{  Nuclear  Physics  Division,  Bhabha  Atomic Research Centre,  Trombay, Mumbai 400 085, India}
\begin{abstract}{ Variational  method  has  been  applied  to estimate   Landau-Pomeranchuk-Migdal (LPM) effects on virtual photon emission 
from the quark gluon plasma as a function of photon mass. The variational method was  well tested for the LPM effects  in  real  photon  
emission.  For virtual photons,  LPM effects  arising  from  multiple  scatterings  of  quarks in the plasma are determined by the integral equations for  the  
transverse   vector  function  (${\bf   \tilde{f}(\tilde{p}_\perp   )}$)  and  the  longitudinal  function ($\tilde{g}({\bf \tilde{p}_\perp } )$).  
We  extended  the  variational  method to solve these  transverse and longitudinal equations for a  variable  set \{$p_0,q_0,Q^2$\},  considering 
bremsstrahlung  and   $\bf  aws$  processes.   We solved these equations, also  by the self consistent  iterations for comparing with the results of  
variational method.  In order to estimate the variational parameter, we obtained empirical fits for the peak positions of the  
${\bf   p_\perp\cdot\tilde{f}(\tilde{p}_\perp   )}$,  
$p_\perp\tilde{g}({\bf \tilde{p}_\perp } )$  distributions from  iteration method. We propose  that the optimized variational parameter 
for virtual photon emission is approximately equal to these  empirical peak position values. The  detailed study  showed that the variational method gives reliable results  for 
LPM effects   on virtual photon emission  for photon virtuality of the order  $Q^2/T^2\le 100$.  At  low $Q^2$,  the peak positions for  $\tilde{p}_\perp$  
distributions of   transverse vector functions for virtual photons nearly coincide  with the peak positions of corresponding  distributions of real photons.  
We calculated imaginary part of photon retarded polarization tensor as a function of   $Q^2/T^2$  using   empirical variational parameters.
}\end{abstract}
\noindent
\maketitle
\par
Study of the physical processes in quark  matter  as  compared  to that  in  hadronic  matter  plays  a  crucial  role  for  identifying  the Quark  gluon  plasma (QGP)  
state,  expected  to  be  formed  in  the  relativistic heavy ion collisions. In this context,   electromagnetic  processes  such  as  photons  and  dilepton  emission are 
important as signals \cite{peitz,gale,rapp} to  identify this de-confined state. In  depth  study  of  photon  emission  processes   in quark-gluon  plasma were presented \cite{kapusta,bair}. 
In the  hard thermal loops \cite{braaten} (HTL) effective theory, Compton scattering  and quark-antiquark  annihilation  processes contribute  at one loop level to  
photon  emission in  quark matter. At  the  two  loop  level,  the  processes  of   bremsstrahlung \cite{brem}  and a crossed process of annihilation  with  scattering  
called ${\bf  aws}$  \cite{bremaws,ktkl} contribute to  photon  emission. Importantly,  these  two  loop  processes contribute  at  the  leading  order  
$O(\alpha\alpha_s)$  resulting  from a special effect called the collinear singularity that is regularized by  the effective  thermal  masses. Owing to the same reasons, 
higher loop multiple scatterings having a ladder topology also contribute at  the same  order  \cite{arnold1,arnold2} giving  a decoherent correction  to  the  two loop 
processes. These rescatterings have been resummed  \cite{arnold1,arnold2}, effectively   implementing   the  Landau-Pomeranchuk-Migdal   (LPM)   effects 
\cite{landau1,landau2,migdal}.  LPM effects  arise due to rescattering of quarks in the medium  during   photon formation  time.  The rescattering  
corrections strongly modify the two loop contributions for bremsstrahlung and $\bf aws$ processes for real photon emission \cite{arnold1,arnold2}.

Photon production  rates from bremsstrahlung and   $\bf  aws$  processes including  the  LPM  effects  are  estimated  by  using  Eq.\ref{photrate}  in terms 
of a transverse  vector  function ${\bf \tilde{p}_\perp \cdot\Re\tilde{f}(\tilde{p}_\perp)}$  \cite{arnold2}.  The resummation of multiple scatterings leads to the 
AMY integral equation for the transverse  vector  function for  real photons,  given in Eq.\ref{amy} \cite{arnold2}.
\begin{eqnarray}
{\cal R}_{b,a}&=& \frac{80\pi T^3\alpha\alpha_s}{(2\pi)^3 9\kappa}\int dp_0
\left[\frac{p_0^2+(p_0+q_0)^2)}{p_0^2(p_0+q_0)^2)}\right]   \times \nonumber \\
&&\left[n_f(q_0+p_0)  (1-n_f(p_0))\right]    \times \nonumber \\
&& ~\int \frac{\bf d^2\tilde{p}_\perp}{(2\pi)^2}  2{\bf \tilde{p}_\perp \cdot\Re\tilde{f}(\tilde{p}_\perp)} \label{photrate} \\
2{\bf \tilde{p}_\perp}&=& i\delta \tilde{E}({\bf \tilde{p}_\perp}) {\bf \tilde{f}}({\bf \tilde{p}_\perp}) +  \nonumber \\
&& \int\frac{d^2{\bf \tilde{\ell}_\perp}}{(2\pi)^2}\tilde{C}({\bf \tilde{\ell}_\perp})
\left[{\bf \tilde{f}}({\bf \tilde{p}_\perp})
-{\bf \tilde{f}}({\bf \tilde{p}_\perp+\tilde{\ell}_\perp})\right]    \label{amy}  \\
\delta \tilde{E}({\bf \tilde{p}_\perp})&=&
\frac{q_0T}{2p_0(q_0+p_0)}\left[\tilde{p}_\perp^2+\kappa \right]   \label{amydelta}
\end{eqnarray}
\begin{eqnarray}
{\cal R}_{b,a}&=&{\cal C}_k \int dp_0
\left[p_0^2+(p_0+q_0)^2)\right]\left[n_f(q_0+p_0 ) \right. \nonumber \\
&& \left. (1-n_f(p_0 ))\right]C_g g(x) \label{gxprc} \\
g(x)&=&g(p_0,q_0,T) \label{gxdef}\nonumber \\
x&=&\frac{1}{\kappa_0}\frac{q_0T}{p_0(p_0+q_0)} \label{xdef} \\
{\cal C}_k&=&\frac{40\alpha\alpha_sT}{9\pi^4q_0^2} \label{ckfact}\\
\kappa_0&=&\frac{M_\infty^2}{m_D^2}\approx \frac{1}{4} ~\mbox{and}~C_g=\frac{\kappa_0}{T} \label{kappa0cg}
\end{eqnarray}
The function $\Re{\bf  \tilde{f}(\tilde{p}_\perp  )}$  in  Eq.\ref{photrate}, which consists of the LPM effects, can be taken as transverse momentum vector
${\bf (\tilde{p}_\perp )}$ times a scalar function of  transverse  momentum ${\tilde{p}_\perp  }$.  The  tilde sign  ~$\tilde{}$~  represents dimensionless 
quantities in units of Debye mass $m_D$ as defined in  \cite{arnold2}.  The function   ${\bf  \tilde{p}_\perp\cdot\Re\tilde{f}(\tilde{p}_\perp  )}$  is determined 
by the AMY  equation (Eq.\ref{amy})  in terms of energy denominator $\tilde{\delta E}$  given in Eq.\ref{amydelta} and the collision kernel 
($\tilde{C}({\bf \tilde{\ell}_\perp})$). We reported that the complex LPM effects can be very well reproduced by introducing the photon emission  function  
$g(x)$ defined in Eq.\ref{gxprc},  of   a dynamical  variable  $x$  defined in Eq.\ref{xdef} \cite{svsprc}.  In  terms  of a single variable function $g(x)$, the 
photon   emission  rates  are   estimated  using Eqs.\ref{gxprc}-\ref{kappa0cg} for any quark momentum, photon energy and  plasma temperature. 
\section{Variational  Method  for Emission of Real Photons} 
AMY integral equation consisting of  LPM effects  was solved using  the variational  approach  and  the photon emission rates were reported in \cite{arnold2}.
In the  present  work, we  follow  this  variational  method and expand the real and imaginary parts of   ${\bf \tilde{f}(\tilde{p}_\perp )}$   in terms of  a  basis   
set   $\{\phi_j\}$ of  trial functions as  shown   in Eqs.\ref{refexpansion}-\ref{basisseti}. The dimensions of the spaces for real and imaginary 
parts  are $N_r$ and $N_i$.  The  energy  function   $\tilde{\delta  E}_{mn}$   and  the  quantities $\tilde{S}_{m,T}$  are calculated by the overlap integrals 
with the  basis trial  functions  as  shown in Eqs.\ref{deltaemnphi},\ref{smntphi}.  The overlap  integrals $\tilde{C}^r_{mn}$ that involve the  collision  kernels 
are shown in  Eq.\ref{cmnphi}.  Another  quantity $\tilde{C}^i_{mn}$  involving  imaginary part  is similar to $\tilde{C}^r_{mn}$.
\begin{eqnarray}
\Re{\bf \tilde{f}(\tilde{p}}_\perp) &=& {\bf \tilde{p}}_\perp \sum_{j=1}^{N_r} a_{j,T} \phi_j^r(\tilde{p}_\perp^2) \label{refexpansion}\\
\Im{\bf \tilde{f}(\tilde{p}}_\perp) &=& {\bf \tilde{p}}_\perp\sum_{j=1}^{N_i} b_{j,T} \phi_j^i(\tilde{p}_\perp^2)   \label{imfexpansion}\\
\phi_j^r(\tilde{p}_\perp^2) &=& \frac{(\tilde{p}_\perp^2/A)^{j-1}}{(1+(\tilde{p}_\perp^2/A)^{N_r+2}} , ~~~~j=1,...,N_r   \label{basissetr} \\
\phi_j^i(\tilde{p}_\perp^2) &=& \frac{(\tilde{p}_\perp^2/A)^{j-1}}{(1+(\tilde{p}_\perp^2/A)^{N_i}}, ~~~~~~ j=1,...,N_i  \label{basisseti}  \\
\tilde{\delta E}_{mn}&=&\left(\phi_m^r,{\tilde{\delta E }}\phi_n^i \right) \label{deltaemnphi}\\
&=& \int \frac{d^2{\bf{\tilde{ p}}}_\perp}{(2\pi)^2} \phi_m^r({\bf{\tilde{ p}}}_\perp){\tilde{ \delta E}} \phi_n^i({\bf {\tilde{p}}}_\perp) \nonumber  \\
\tilde{S}_{m,T} &=& \left(2{\bf {\tilde{p}}}_\perp, {\bf {\tilde{p}}}_\perp\phi_m^r\right)  \label{smntphi}\\
\tilde{C}^r_{mn} &=& \frac{1}{2}\int \frac{d^2{\bf \tilde{p}}_\perp}{(2\pi)^2}\frac{d^2{\bf \tilde{q}}_\perp}{(2\pi)^2}
\tilde{C}({\bf \tilde{q}}_\perp)  \left[{\bf \tilde{p}}_\perp\phi_m^r({\bf \tilde{p}}^2) \right. \nonumber \\
&& -\left. ({\bf \tilde{p}}_\perp+{\bf \tilde{q}}_\perp)\phi_m^r( |{\bf \tilde{p}}_\perp+{\bf \tilde{q}}_\perp|^2   ) \right]
\cdot\left[ {\bf \tilde{p}}_\perp\phi_n^r({\bf \tilde{p}}^2) \right. \nonumber \\
&& -\left. ({\bf \tilde{p}}_\perp+{\bf \tilde{q}}_\perp)\phi_n^r( |{\bf \tilde{p}}_\perp+{\bf \tilde{q}}_\perp|^2   ) \right] \label{cmnphi}
\end{eqnarray}
$a_{j,T},b_{j,T}  $  in the  Eqs.\ref{refexpansion},\ref{imfexpansion} are  the expansion coefficients for real and imaginary parts of  ${\bf \tilde{f}(\tilde{p}}_\perp) $
respectively.  Here, the  subscripts  T and also in $\tilde{S}_{m,T}$ represent transverse vector function ${\bf \tilde{f}}{\bf (\tilde{p}_\perp )}$.  Using the trial 
functions, some  of these  quantities can  be  evaluated  analytically or simplified as shown in Eqs.\ref{deltaemn}-\ref{cmn}. The choice of dimensions for real and 
imaginary parts should be pragmatically large enough  for performing calculations. For real photon emission calculations, we  used  two flavors, 
three colors, $\alpha_s$=0.20 and   $N_r=N_i=8$. A few cases were verified for convergence  using  $N_r=N_i$=10 and 12. These results for photon emission
rates  were reported earlier \cite{svsprc}. However, it is instructive to present  the details of variational calculations and analysis of results for the case of real photons.
\begin{eqnarray}
\tilde{\delta E}_{m,n,T}&=&\frac{q_0TA^2}{2p_0r_0} K_T  \label{deltaemn} \\
K_T&=& A K(m+n,N) + \kappa K(m+n-1,N) \nonumber \\
\tilde{S}_{m,T}&=&2A^2K(m,N_r)   \label{smn} \\
K(m,n) &=& \frac{m!(N-m)!}{4\pi(N+1)!}   \label{kmn} \\
N &=& N_r + N_i \label{nrni} 
\end{eqnarray}
\begin{eqnarray}
\tilde{C}^r_{mn} &=& \frac{1}{32\pi^2}\int d\tilde{p}_\perp^2d\tilde{q}_\perp^2\int_{-\pi}^\pi\frac{d\theta}{2\pi} \tilde{C}(\tilde{q}_\perp) \nonumber \\
&& \left\{\tilde{p}_\perp^2\phi_m^r(\tilde{p}_\perp^2)\phi_n^r(\tilde{p}_\perp^2)+\right.\nonumber \\
&&|\tilde{\bf{p}}_\perp+\tilde{\bf{q}}_\perp|^2\phi_m^r(|\tilde{\bf p}_\perp+\tilde{\bf q}_\perp|^2) \phi_n^r(|\tilde{\bf p}_\perp+\tilde{\bf q}_\perp|^2)\nonumber \\
&&-\left( \tilde{p}_\perp^2+\tilde{p}_\perp\tilde{q}_\perp\cos\theta\right)\left[\phi_m^r(|\tilde{\bf p}_\perp+\tilde{\bf q}_\perp|^2) \phi_n^r(\tilde{p}_\perp^2) \right.\nonumber \\
&&  \left. \left.  + (m \leftrightarrow n)  \right] \right\}   \label{cmn}
\end{eqnarray}
In the  variational  approach,  the trial functions consist of a variational parameter $A$  which needs to be optimized to obtain correct  results. The    
${\bf  \tilde{p}_\perp\cdot\Re\tilde{f}(\tilde{p}_\perp)}$  distributions  are  usually sharply peaked  and  therefore  the  optimized  value  of the variational 
parameter $A_v (A=A_v^2)$  should be around the peak positions of these  distributions.  In our earlier paper, we simplified this variational approach and extended this method
 to  finite baryon density case \cite{svs1}. It was shown  that   the  variational  parameter can be taken as  $A =A_v^2(p_0,q_0,T)$,  together with  
$A_v(p_0,q_0)=\left|{p_0(q_0+p_0)}/{q_0}\right|^{0.31}$ and constraint  $A_v\ge0.31$.   In the calculations that follow, we  used  this  empirical   result  for  the variational  parameter.   In  
order  to  test  the empirical  values of the  variational  parameter, we  transform  the  inner  product  integral ${\bf \tilde{p}_\perp \cdot\Re\tilde{f}(\tilde{p}_\perp)}$ 
defined  over [0,$\infty$]  to a finite  interval  of  [0,1]  by  defining  a variable  $y$  in   Eq.\ref{yfunc} (see \cite{arnold2}).  In  terms  of  $y$, this inner product 
 is given in Eq.\ref{yintegrand}.  Then we  estimate  the  value  of ${\tilde{p}_\perp}$  where  the  integrand  in Eq.\ref{yintegrand} peaks. If the value of the 
variational  parameter $A(=A_v^2)$ used is  correct, the result should be  $y=\frac{1}{2}$  at  this  peak position ${\tilde{p}_\perp}$.  Therefore, finding the optimized value of
the variational parameter is equivalent to finding the peak positions of these ${\bf  \tilde{p}_\perp\cdot\Re\tilde{f}(\tilde{p}_\perp)}$ distributions.
\begin{eqnarray}
y&=&\frac{|{\bf \tilde{p}}_\perp|}{A^{1/2}+|{\bf \tilde{p}}_\perp|}.  \label{yfunc} \\
\left({\bf \tilde{p}}_\perp,{\bf \tilde{f}}\right) &=& \frac{1}{\pi}\int_0^1 \frac{A^2y^3}{(1-y)^5}\chi\left(Ay^3/(1-y)^2\right)dy.
\label{yintegrand}
\end{eqnarray}
\begin{figure}[!]
\hspace{-.50cm}
\includegraphics[height=14.cm,width=9.25cm]{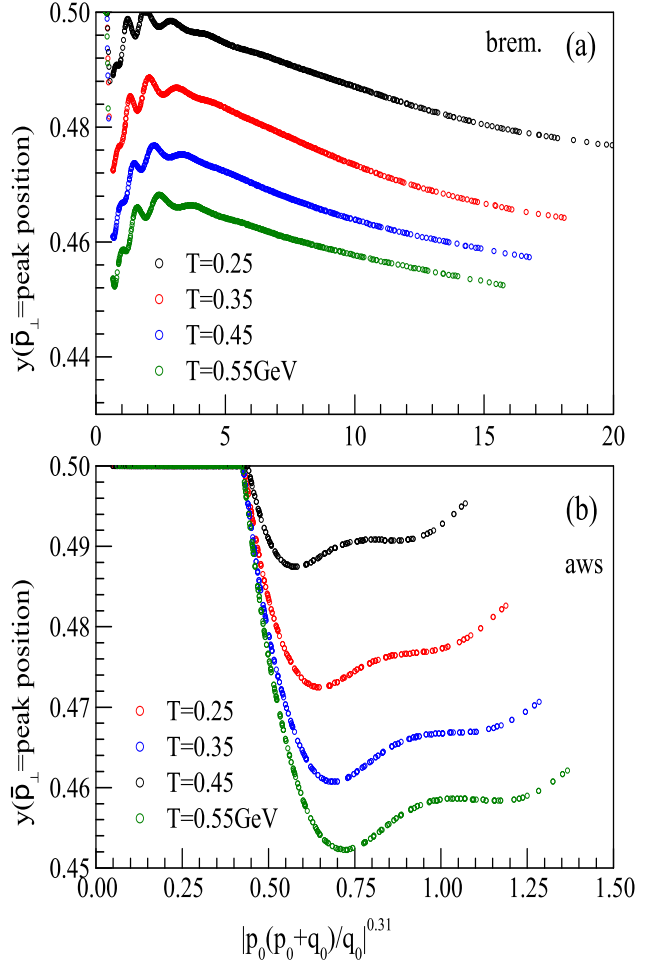}
\caption{  The  $y$  function  constructed  from the peak positions of real photon ${\bf \tilde{p}_\perp}\cdot \Re{\bf  \tilde{f}(\tilde{p}_\perp  )}$ distributions.
Symbols  represent the peak positions of ${\bf p_\perp}$ distributions from variational method for different  $\{p_0,q_0,T\}$  values.  The
variational  method  used  an  empirical  formula for variational parameter. The $y$ value should be 1/2 for  truly  optimized  variational
parameter.  Figure  shows  that  the empirical variational parameter values used for real photons  satisfied this constraint approximately.}
\label{realyba}
\end{figure}
\begin{figure}[!]
\hspace{-1.2cm}
\includegraphics[height=14.cm,width=9.25cm]{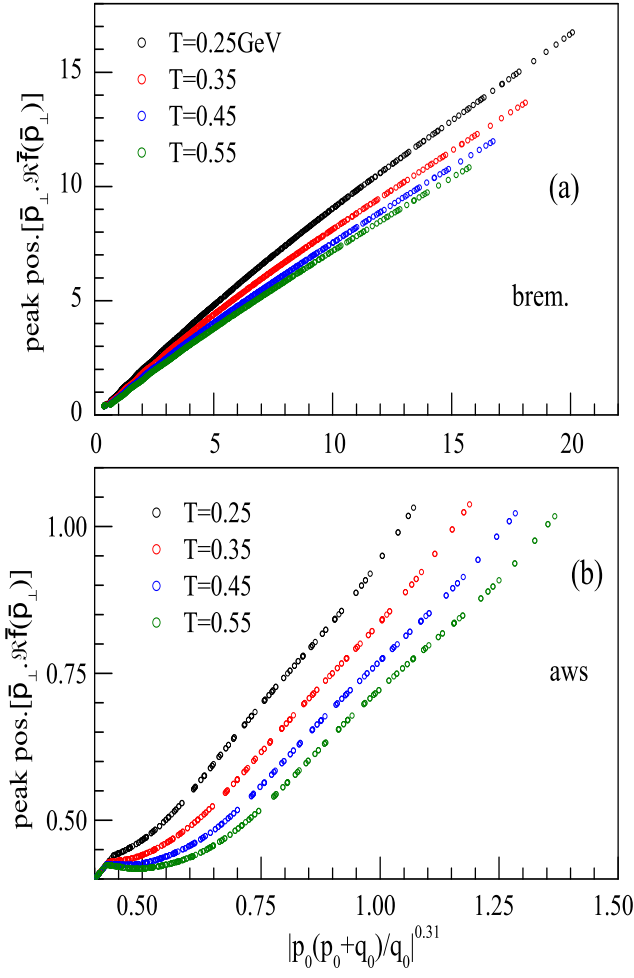}
\caption{  The  peak  positions  of real photon ${\bf \tilde{p}_\perp}\cdot \Re{\bf \tilde{f}(\tilde{p}_\perp )}$  distributions versus (peak positions from)  
empirical formula.  Symbols represent the peak positions  from  variational  method  for different  $\{p_0,q_0,T\}$  values.}
\label{realppba}
\end{figure}
\begin{figure}[!]
\hspace{-.75cm}
\includegraphics[height=20.cm,width=9.25cm]{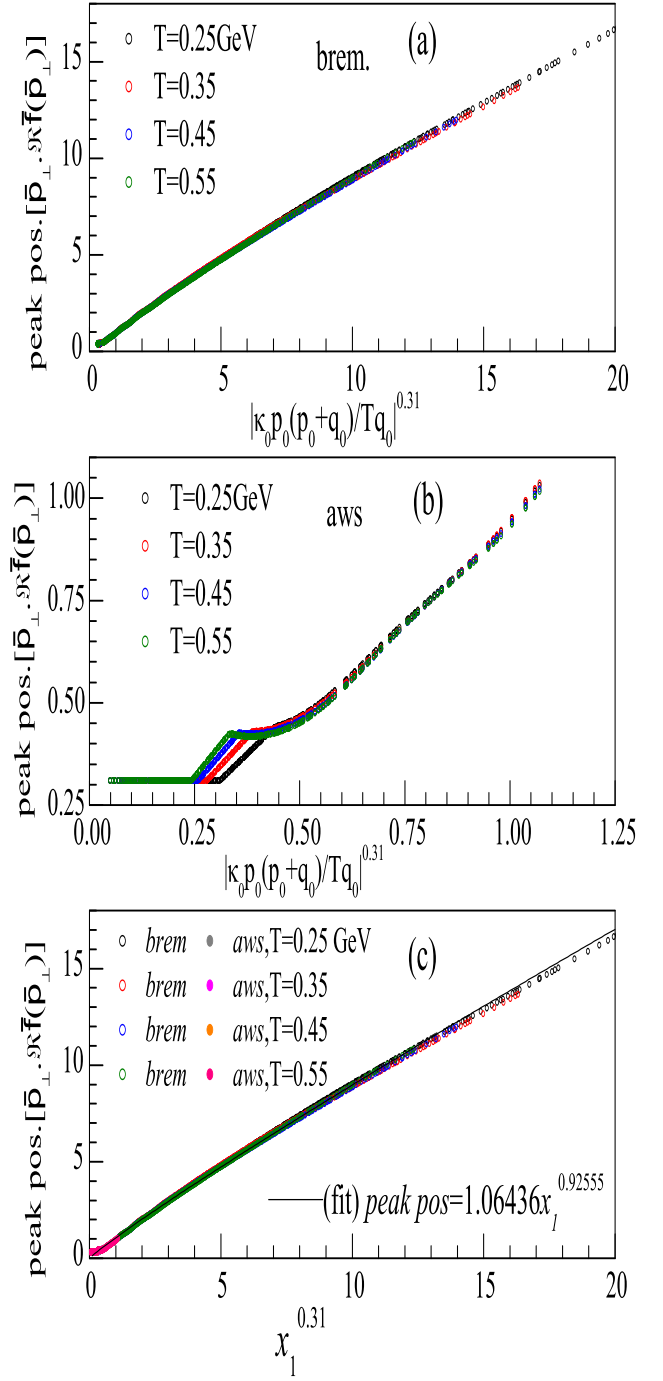}
\caption{The  peak  positions  of real photon ${\bf \tilde{p}_\perp}\cdot \Re{\bf \tilde{f}(\tilde{p}_\perp )}$  distributions versus (peak positions from)  empirical
formula  values, for (a)  bremsstrahlung  and for (b) ${\bf aws}$.  Symbols represent the peak positions of ${\bf p_\perp}$  distributions  from  variational  method  for different  
$\{p_0,q_0,T\}$  values.  Figure  shows  that the $x_1$ variable defined in text  (Eq.\ref{x1})  for  real  photons  is  a  good  scale. The data of (a) and (b) are all
plotted in figure (c). The empirical fit function and its parameters are shown in figure (c) which can be used to fix the optimized variational parameter.}
\label{realppxtallba}
\end{figure}
We observed that these distributions are not very sensitive to the  exact  value  of  the  $A_v$ parameter   provided  the    dimension   is   sufficiently   high.   
Figure. \ref{realyba} shows  $y$ at the  peak  positions  of  these  ${\bf \tilde{p}_\perp \cdot\Re\tilde{f}(\tilde{p}_\perp)}$ distributions.  
Figure.    \ref{realyba}(a)    shows   $y$   for   bremsstrahlung   and \ref{realyba}(b)   for   ${\bf aws}$   processes,   as    a    function    of 
$\left|{p_0(q_0+p_0)}/{q_0}\right|^{0.31}$.    The    figure   includes variational   calculations   for   a   set   of   2304  values of  $\{p_0,q_0,T\}$  (=24x24x4   
values {\it  i.e.,}  24  for $p_0$, 24 for  $q_0$ and 4  temperatures). It can be seen that the $y$ value is  close  to $\frac{1}{2}$ for 
both bremsstrahlung and {\bf aws} processes. However deviation of $y$ from 1/2  increases  with  increasing  temperature, as temperature dependence for 
variational parameter was  not  considered.  The exact   peak   position   values   of  these  distributions  are  shown  in Figure.\ref{realppba}(a) 
for bremsstrahlung and \ref{realppba}(b) for       ${\bf aws}$  processes,   as  a  function  of $\left|{p_0(q_0+p_0)}/{q_0}\right|^{0.31}$.  As  
shown  in  figure, the peak positions are linear  and  the  temperature  dependence  is  not  very strong.   However,   when   these   peak   positions   are  
plotted  versus $x_1^{0.31}$ with  $x_1=\left|\kappa_0p_0(q_0+p_0)/(Tq_0)\right|$, all the 2304 points of $\{p_0,q_0,T\}$ merge for bremsstrahlung process 
and similarly for ${\bf aws}$ process as shown in Figs.\ref{realppxtallba}(a,b).  Further, when the bremsstrahlung and {\bf aws} peak positions are both together 
plotted versus $x_1^{0.31}$, all the data points of Fig.\ref{realppxtallba}(a,b) overlap as shown in Fig.\ref{realppxtallba}(c). Therefore, the peak 
positions  of these two processes follow   scaling with $x_1$ variable. This is not surprising because we have shown earlier that $x_1$ is an exact 
dynamical  scale for real  photons\cite{svsprc}.  We have fitted the data by a suitable function and the parameters are shown in Fig.\ref{realppxtallba}(c). Therefore for real 
photon emission, the optimized variational parameter  is approximately  given by the empirical fit values shown in Fig.\ref{realppxtallba}(c)  for any temperature, quark 
momentum and photon energy values. 
\section{Variational  Method  for  virtual Photon emission}   
Processes  that contribute  to  virtual  photon  emission  in QGP at $\alpha\alpha_s$ order \cite{alther} and the higher order corrections \cite{thoma} were  
reported.  The HTL one loop   processes   $q\bar{q}\rightarrow   g\gamma^*$,  and  $qg\rightarrow q\gamma^*$ contribute  to  photon  polarization  tensor
at  the  order   $\alpha\alpha_s$.  Photon emission from QGP as a function  of photon  mass  considering LPM effects  was  also  reported  \cite{lpmdilep}. 
It was shown that the multiple scatterings  modify  the  imaginary  part  of  self energy as a function of photon energy and  momentum both, especially modifying 
 the   tree  level  threshold,  $Q^2  =  4M_\infty^2$ \cite{lpmdilep}. The dilepton emission rates are estimated in terms  of  the imaginary   part   of   retarded   
photon  polarization  tensor for virtual photons,  given  by \cite{lpmdilep}.
\begin{equation}
\frac{dN_{\ell^+\ell^-}}{d^4xd^4Q} = \frac{\alpha_{EM}}{12\pi^4Q^2(e^{q0/T}-1)} \Im\Pi_{R\mu}^\mu(Q)
\end{equation}
$\Im\Pi_{R\mu}^\mu(Q)$ is determined by transverse and longitudinal functions represented by ${\bf f(p_\perp)},g(p_\perp)$ resulting from two integral equations.  
For  the case of  virtual  photon  emission, the  transverse vector function is determined by  the  Eq.\ref{agmz-t} and  the energy   transfer 
$\delta  {E}(({\bf  {p}_\perp},p_0,q_0,Q^2)$ \cite{lpmdilep}. For  virtual  photons,  the  coupling  of quarks  to  longitudinal  mode must be considered. 
This results in a scalar function of ${\bf p_\perp}$ which is determined by the AGMZ  integral equation  given in Eq.\ref{agmz-l}  with the collision kernel  
from \cite{kernel}.  For the case  of  massive  photon  emission,  this  energy  denominator  $\delta{E}$    is modified  from that  of real photons by    
replacing     $M_\infty^2        \rightarrow M_{\mbox{eff}}=M_\infty^2+\frac{Q^2}{q_0^2}{p_0r_0}$ as  in Eq.\ref{agmz-de}.  For  
$Q^2>4M_\infty^2$,    this   M$_{\mbox{eff}}$  can  vanish  or  even  become negative.  $\delta{E}$ energy denominator  can be interpreted as 
inverse  formation time of the photon, which acquires dependence on photon mass  in addition to the dependence on  photon  energy, quark momentum 
of a real photon. 
\begin{eqnarray}
2{\bf {p}_\perp}&=&i\delta {E}({\bf {p}_\perp},p_0,q_0,Q^2) {\bf {f}}({\bf {p}_\perp}) +  g^2C_FT \nonumber \\
&& \int \frac{d^2{\bf{\ell}_\perp}}{(2\pi)^2}{C}({\bf {\ell}_\perp}) \left[{\bf {f}}({\bf {p}_\perp})
-{\bf {f}}({\bf {p}_\perp+{\bf\ell}_\perp})\right]    \label{agmz-t}  \\
2{\bf \sqrt{|p_0r_0|}}&=&i\delta {E}({\bf{p}_\perp},p_0,q_0,Q^2)
g({\bf {p}_\perp}) + g^2C_FT   \nonumber \\
&&\int \frac{d^2{\bf {\bf\ell}_\perp}}{(2\pi)^2}C({\bf {\ell}_\perp})\left[g({\bf {p}_\perp})-g({\bf{p}_\perp+\bf{\ell}_\perp})\right]~~~~~\label{agmz-l}
\end{eqnarray}
\begin{eqnarray}
\delta E({\bf{p}_\perp},p_0,q_0,Q^2)&=&\frac{q_0}{2p_0r_0}\left[{\bf{p}_\perp^2}+M_{\mbox{eff}}^2 \right]     \label{agmz-de}
\end{eqnarray}
In  above  equations, $r_0=p_0+q_0$ and ${\bf{f}},g$ are actually functions of $p_0,q_0,Q^2$ represented as, ${\bf {f}}({\bf  {p}_\perp},p_0,q_0,Q^2)$,
$ g({\bf {p}_\perp},p_0,q_0,Q^2)$.  Eq.\ref{agmz-t} and Eq.\ref{amy} are identical except for $\delta E$ energy factor. Further, Eq.\ref{agmz-t} and the  
Eq.\ref{agmz-l}  are  similar on the right side of the equations, however  the left side of AGMZ equation is a constant $\sqrt{|p_0r_0|}$. Aurenche {\it  et.  al.,}  
solved these  equations,  based  on  a method  of impact parameter representation \cite{lpmdilep}. As shown in previous section, we  have  solved  the AMY  
equation  for real photons by the variational method and a new method called iterations method by formulating these equations   in  terms  of  tilded  variables. 
Therefore,  we will  transform the two above equations Eqs.\ref{agmz-t},\ref{agmz-l}   to tilded quantities, and for details see \cite{arnold2,gef}.  The 
transformation for  ${\bf p_\perp, f(p_\perp)},\delta E({\bf p_\perp}),C({\bf p_\perp})$ are given by Eq.\ref{tildetrans}. 
\begin{eqnarray}
{\bf    \tilde{p}_\perp}&=&\frac{\bf p_\perp}{m_D}~~;~~{\bf \tilde{f}(\tilde{p}_\perp)}=\frac{m_D}{T}{\bf{f}({p}_\perp)} \nonumber \\
{\tilde{\delta E}}({\bf\tilde{p}_\perp})&=&\frac{T}{m_D^2}{\delta E(\bf{p}_\perp}) ~~;~~ \tilde{C}{\bf(\tilde{\bf\ell}_\perp})=T C({\bf\ell_\perp}). \label{tildetrans}
\end{eqnarray}
\begin{eqnarray}
2{\bf \tilde{p}_\perp}&=& i \delta \tilde{E}({\bf \tilde{p}_\perp},p_0,q_0,Q^2)
{\bf \tilde{f}}({\bf \tilde{p}_\perp}) \nonumber \\
&& + \int\frac{d^2{\bf \tilde{\bf\ell}_\perp}}{(2\pi)^2} \tilde{C}({\bf \tilde{\ell}_\perp})
\left[{\bf \tilde{f}}({\bf \tilde{p}_\perp}) -{\bf \tilde{f}}({\bf \tilde{p}_\perp+\tilde{\bf\ell}_\perp})\right]   \label{agmz-t-tilde}
\end{eqnarray}
\begin{equation}
\delta \tilde{E}({\bf \tilde{p}_\perp},p_0,q_0,Q^2)= \frac{q_0T}{2p_0(q_0+p_0)}\left[\tilde{p}_\perp^2+\kappa_{\mbox{\small{eff}}}\right]
\end{equation}
In    above $\kappa_{\small{\mbox{eff}}}=\frac{M^2_{\mbox{\small{eff}}}}{m_D^2}$.    We will divide the  above  Eq.\ref{agmz-l},  by  $m_D$  in  order 
to get the following equation, where absorbing 1/$m_D$ factor, $g$ is now re-defined.  The equation for the longitudinal part could be written as,
\noindent
\begin{eqnarray}
2\sqrt{\frac{|p_0r_0|}{m_D^2}}&=& i\delta \tilde{E}({\bf \tilde{p}_\perp},p_0,q_0,Q^2)
{\tilde{g}}({\bf \tilde{p}_\perp}) + \nonumber \\
&& \int\frac{d^2{\bf \tilde{\ell}_\perp}}{(2\pi)^2}\tilde{C}({\bf \tilde{\ell}_\perp}) \left[{\tilde{g}}({\bf \tilde{p}_\perp}) -{\tilde{g}}({\bf \tilde{p}_\perp+\tilde{\bf\ell}_\perp})\right] ~~~~~
\label{agmz-l-tilde-mod}
\end{eqnarray}
In the above equation,  $g$ transforms as $\tilde g({\bf \tilde{p}_\perp})= \frac{m_D}{T} g({\bf p_\perp})$, similar to $\bf f(p_\perp)$ function.  It  implies  
that the $\tilde{g}$ is larger than $g$ by a factor $m_D$.  Therefore, when the solution for this Eq.\ref{agmz-l-tilde-mod} is obtained and  integrated  over  
$\int\frac{d^2\bf \tilde{p}_\perp}{(2\pi)^2}\Re{\tilde{g}{\bf    (\tilde{p}_\perp)}}$,    the  result will be larger than the true result from Eq.\ref{agmz-l}
by exactly $m_D$ factor \cite{gef}. In  the present work, we have solved the above Eq.\ref{agmz-l-tilde-mod} by iterations and variational method with 
an aim to test the validity of  the variational  method  for a virtual photon case. The variational method for  longitudinal part  has  been  re-derived  as shown 
in Eqs.\ref{regexpansion}-\ref{cmnrlvirt}  \cite{svssymp}.  The basis for expansion of the ${\bf \tilde{f}(\tilde{p}}_\perp)$ is same as in Eqs.\ref{refexpansion}-\ref{basisseti}.
The equations for $\tilde{\delta E}_{mn}$ and  $\tilde{S}_{m,T}$ remain same as in Eqs.\ref{deltaemn},\ref{smn} except for the 
change $\kappa \rightarrow \kappa_{\mbox{eff}}$.
\begin{eqnarray}
\Re{\tilde{g}(\bf \tilde{p}}_\perp) &=&  \sum_{j=1}^{N_r} a_{j,L} \phi_j^r(\tilde{p}_\perp^2)  \label{regexpansion}\\
\Im{\tilde{g}(\bf \tilde{p}}_\perp) &=& \sum_{j=1}^{N_i} b_{j,L} \phi_j^i(\tilde{p}_\perp^2)  \label{imgexpansion}\\
\phi_j^r(\tilde{p}_\perp^2) &=& \frac{(\tilde{p}_\perp^2/A)^{j-1}}{(1+\tilde{p}_\perp^2/A)^{N_r+2}} , ~~~~j=1,...,N_r \label{basiselr}\\
\phi_j^i(\tilde{p}_\perp^2) &=& \frac{(\tilde{p}_\perp^2/A)^{j-1}}{(1+\tilde{p}_\perp^2/A)^{N_i}}, ~~~~~~~~ j=1,...,N_i \label{basiseli}
\end{eqnarray}
\begin{eqnarray}
\tilde{\delta E}_{mn}&=&\left(\phi_m^r,{\tilde{\delta E }}\phi_n^i \right) \\
&=& \int \frac{d^2{\bf{\tilde{ p}}}_\perp}{(2\pi)^2} \phi_m^r({\bf{\tilde{ p}}}_\perp){\tilde{ \delta E}} \phi_n^i({\bf {\tilde{p}}}_\perp)  \label{deltaevirt}\nonumber \\
\tilde{\delta E}_{mn,L}&=&\frac{kTA}{2p_0r_0} K_L \label{deltaemnvirt}\\
K_L&=&A K(m+n-1,N)+\kappa_{\small{\mbox{eff}}} K(m+n-2,N)  \label{klvirt} \nonumber \\
\tilde{S}_{m,T} &=& \left(2{\bf {\tilde{p}}}_\perp, {\bf {\tilde{p}}}_\perp\phi_m^r\right) \\
&=&2A^2K(m,N_r)   \label{smtvirt}\\
\tilde{S}_{m,L} &=& \left(2\sqrt{\frac{p_0r_0}{m_D^2}}, \phi_m^r\right)   \\
&=&\sqrt{\frac{p_0r_0}{m_D^2}}K(m-1,N_r)  \label{smlvirt} \\
\tilde{C}^r_{mn,L} &=& \frac{1}{2}\int \frac{d^2{\bf \tilde{p}}_\perp}{(2\pi)^2}\frac{d^2{\bf \tilde{q}}_\perp}{(2\pi)^2} \tilde{C}({\bf \tilde{q}}_\perp)  \left[\phi_m^r({\bf \tilde{p}}^2) \right. \nonumber \\
&& -\left. \phi_m^r( |{\bf \tilde{p}}_\perp+{\bf \tilde{q}}_\perp|^2   ) \right]\cdot\left[ \phi_n^r({\bf \tilde{p}}^2) \right. \nonumber \\
&& -\left. \phi_n^r( |{\bf \tilde{p}}_\perp+{\bf \tilde{q}}_\perp|^2   ) \right]  \label{cmnrphilvirt}\\
K(m,n) &=& \frac{m!(N-m)!}{4\pi(N+1)!}  \nonumber 
\end{eqnarray}
\begin{eqnarray}
\tilde{C}^r_{mn} &=& \frac{1}{32\pi^2}\int d\tilde{p}_\perp^2d\tilde{q}_\perp^2\int_{-\pi}^\pi\frac{d\theta}{2\pi}\tilde{C}(\tilde{q}_\perp)\left\{\phi_m^r(\tilde{p}_\perp^2)\phi_n^r(\tilde{p}_\perp^2)\right.\nonumber \\
              &&+  \phi_m^r(|\tilde{\bf p}_\perp+\tilde{\bf q}_\perp|^2) \phi_n^r(|\tilde{\bf p}_\perp+\tilde{\bf q}_\perp|^2)    \nonumber \\
             &&-  \left[\phi_m^r(|\tilde{\bf p}_\perp+\tilde{\bf q}_\perp|^2) \phi_n^r(\tilde{p}_\perp^2) \right.  \nonumber \\
            &&  \left. \left.  + (m \leftrightarrow n)  \right] \right\} \label{cmnrlvirt}
\end{eqnarray}
The equations for expansion of $\tilde{g}({\bf \tilde{p}}_\perp)$ and the basis functions  are similar to transverse part  and the equations are given 
in Eqs.\ref{regexpansion}-\ref{basiseli}.  The energy denominator and other quantities are shown in Eqs.\ref{deltaevirt}-\ref{cmnrlvirt}.  It is important to note 
that the variational parameter ~ $A$ ~ in Eqs.\ref{regexpansion}-\ref{cmnrlvirt} is in general different for transverse and longitudinal parts. Use of same symbol is misleading 
and further, the variational parameters for these two parts need to be optimized independently. This is because, in addition to overall $p^2_\perp$ factor difference 
in ${\bf \tilde{p}_\perp}\cdot \Re{\bf  \tilde{f}(\tilde{p}_\perp  )}$,~ $\Re\tilde{g}(\tilde{\bf p}_\perp  )$,  the  equations differ also in the interference 
terms in Eqs.\ref{cmn},\ref{cmnrlvirt}. In all the calculations that follow, we have used two flavors, three colors,  $\alpha_s$=0.30, $N_r=N_i=10$.  and 
T=1.0GeV.    Following  the iterations and variational methods, we obtained  the $\bf p_\perp$ distributions for the  bremsstrahlung  and  {\bf aws}  cases  for 
both transverse and longitudinal components. We obtained  350  distributions  (5  for $q_0$, 10 for $Q^2$  and 7 for $p_0$ values)  for  transverse   and  
350 distributions for longitudinal parts of  bremsstrahlung. Similarly the  distributions were obtained for {\bf  aws} case. All these distributions  were 
generated in   iterations method and many of the  cases also by variational method.  The results of these  two methods were  compared. 
The variational parameter has been varied to optimize the distributions. It was  observed that for a given $q_0$, the variational method  does  not give  
satisfactory results for increasing $Q^2$ values and further variational distributions  showed large oscillations.  However, when the $Q^2$ is low, the agreement  
of  results from variational and iterations methods is very good. Therefore, variational  method can be used for low $Q^2$ values very reliably. The iterations  
gave correct  converging results for  all $q_0,Q^2$ values studied. Therefore in the present work, the iterations method is  taken as reference  standard for 
these $\bf p_\perp$ distributions.  These details will be presented  in the next section. 
\section{Empirical  Analysis of the Solutions of AMY, AGMZ integral equations}
\begin{figure}[!]
\hspace{-1.00cm}
\includegraphics[height=20.cm,width=9.5cm]{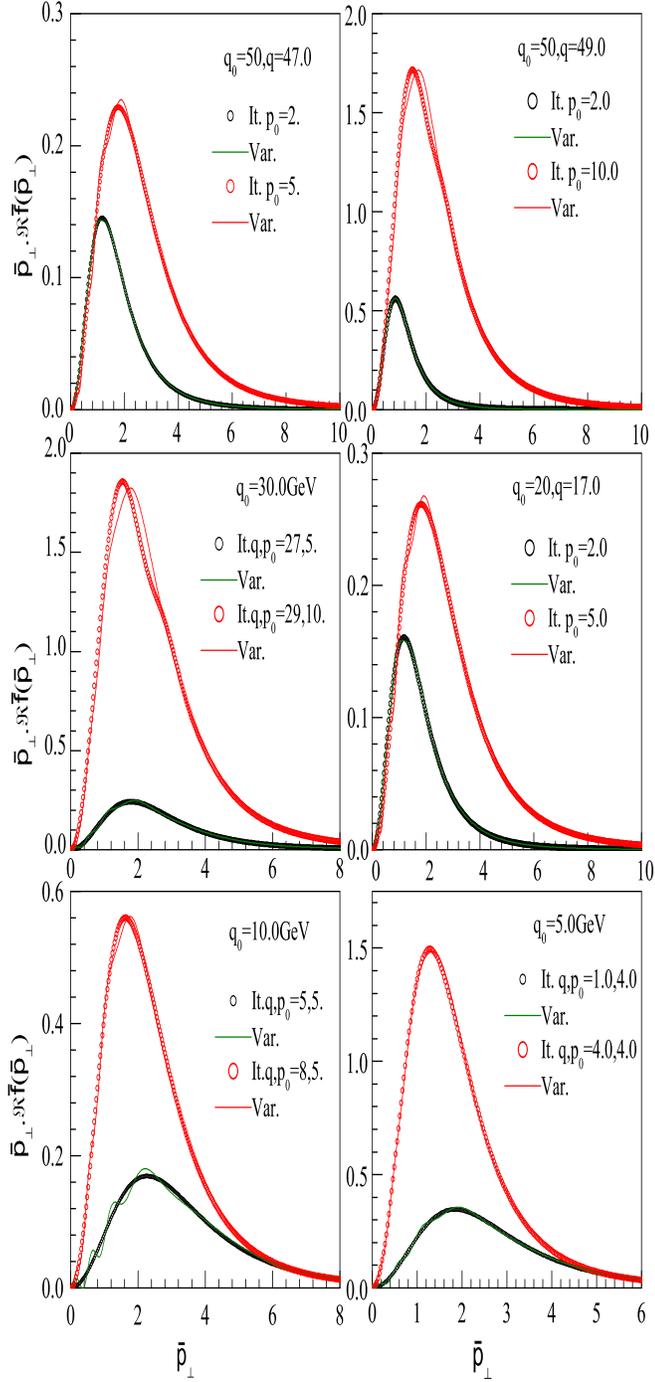}
\caption{  The  ${\bf p_\perp}$   distributions  ${\bf \tilde{p}_\perp} \cdot\Re{\bf \tilde{f}(\tilde{p}_\perp)}$ for  transverse part of bremsstrahlung photon 
emission.  The symbols represent  results of iterations.  About 350 cases of  $\{p_0,q_0,Q^2\}$ values were studied and some are 
shown in the figure and see text for details. Different colored symbols represent different quark momenta.  The results of variational method are shown by 
curves. The terms It. and Var. in figure labels refer to iterations and variational methods. }
\label{virtbt}
\end{figure}

\begin{figure}[!]
\hspace{-1.0cm}
\includegraphics[height=10.cm,width=9.50cm]{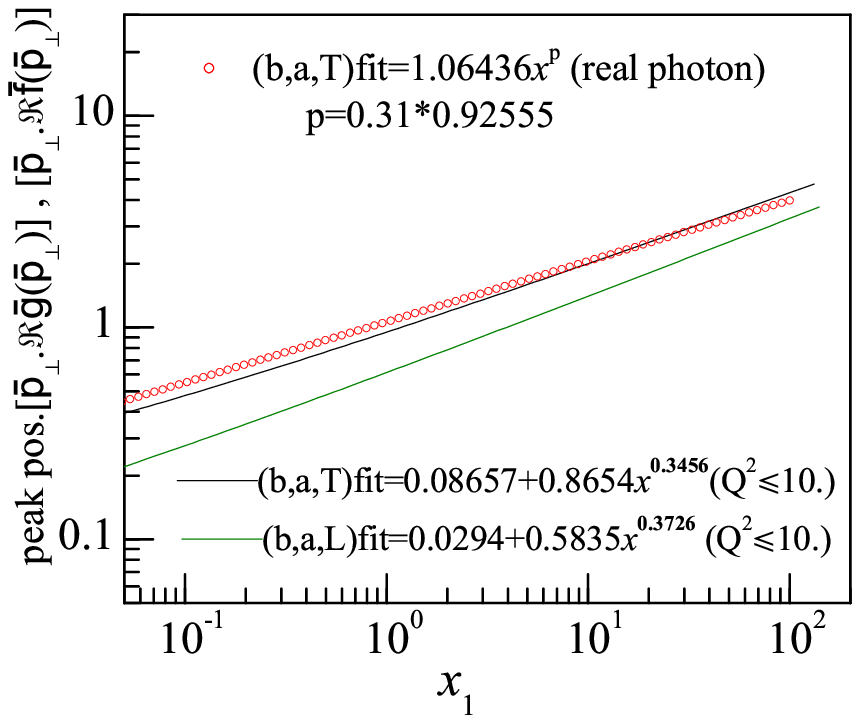}
\caption{ The peak positions of the   ${\bf  \tilde{p}_\perp}\cdot{\bf  \Re\tilde{f}}({\bf  \tilde{p}_\perp})$ for real photons 
represented by red circles, same as in Fig.\ref{realppxtallba}(c). The peak positions  
${\bf  \tilde{p}_\perp}\cdot{\bf  \Re\tilde{f}}({\bf  \tilde{p}_\perp})$,~ ${ \tilde{p}_\perp} \Re\tilde{g}({\bf \tilde{p}_\perp})$ 
distributions  corresponding to various processes  and photon virtuality $Q^2\le10GeV^2$ are shown by black curve and green curve respectively. }
\label{realvirt1}
\end{figure}

\noindent
Figure \ref{virtbt} shows the  ${\bf p_\perp}$   distributions for ${\bf \tilde{p}_\perp} \cdot\Re{\bf \tilde{f}(\tilde{p}_\perp)}$   (transverse part of) 
bremsstrahlung process for five values of $q_0=50,30,20,10,5$ GeV. For each $q_0$, we have shown distributions (in different colors in figure) for two 
different photon  and quark momenta $(Q^2,p_0)$ values. The symbols represent results from iterations method and the curves from  the variational 
method.  In many cases, the curves are not visible as the symbols are overwritten on curves, suggesting that the agreement is very good. 
However, as the $Q^2$ increases, these distributions from the variational method show  oscillations and deviate from the  iterations method. The agreement of 
variational results with iterations is good for all cases of  $Q^2$ less than the higher $Q^2$ shown in the figures. For example, in the  Figure \ref{virtbt}, the 
deviation increases for any $Q^2\ge(50^2-47^2)GeV^2$  for $q_0=50$GeV ~;~ $Q^2\ge(30^2-27^2)GeV^2$  for $q_0=30$GeV ~;~ 
 $Q^2\ge(20^2-17^2)GeV^2$  for $q_0=20$GeV ~;~ $Q^2\ge(10^2-5^2)GeV^2$  for $q_0=10$GeV ~;~ $Q^2\ge(5^2-1^2)GeV^2$  for $q_0=5$GeV ; \\
\begin{figure}[!]
\hspace{-1.0cm}
\includegraphics[height=20.cm,width=9.50cm]{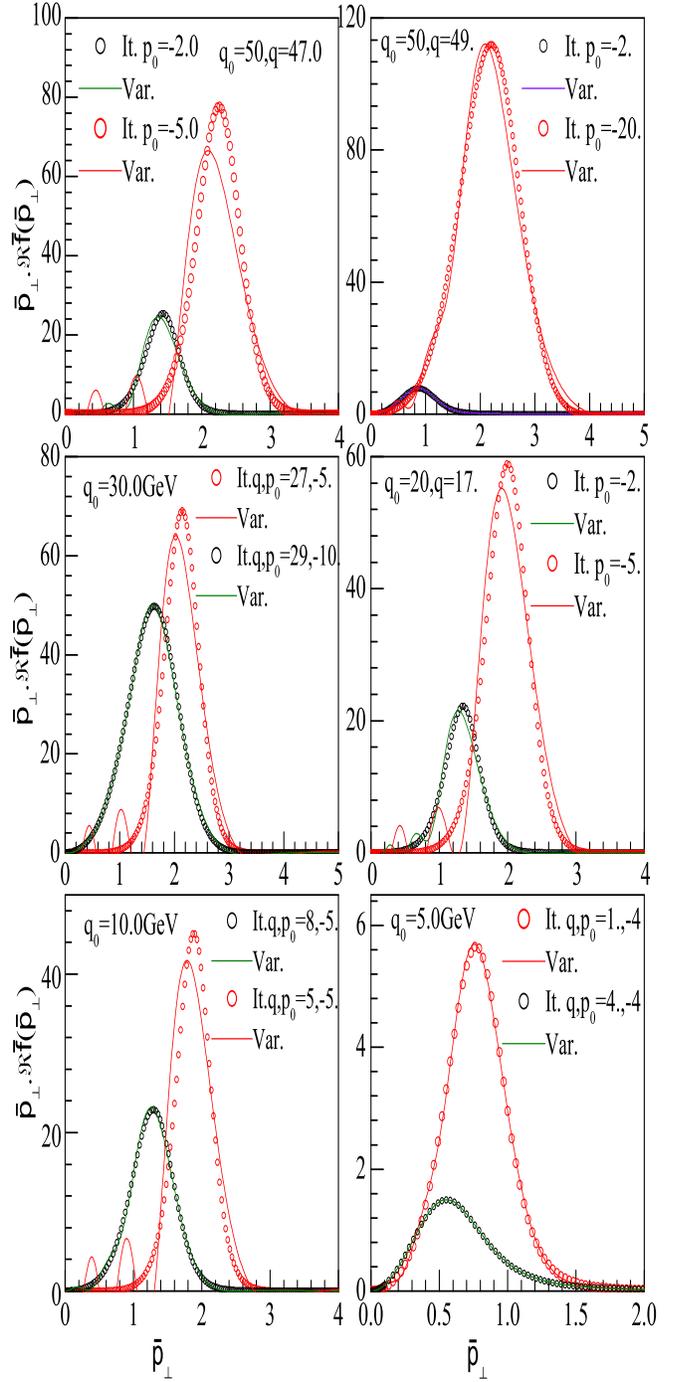}
\caption{   The  ${\bf p_\perp}$   distributions  ${\bf \tilde{p}_\perp}\cdot\Re{\bf \tilde{f}(\tilde{p}_\perp)}$  for  {\bf aws} process. The symbols 
represent the   results from  iterations for some cases of $\{p_0,q_0,Q^2\}$ values studied.   Different colored symbols represent different quark momenta. 
The results of variational method are shown by  curves. The terms It. and Var. in figure labels refer to iterations and variational methods.}
\label{virtat}
\end{figure}
\begin{figure}[!]
\hspace{-1.0cm}
\includegraphics[height=21.cm,width=9.5cm]{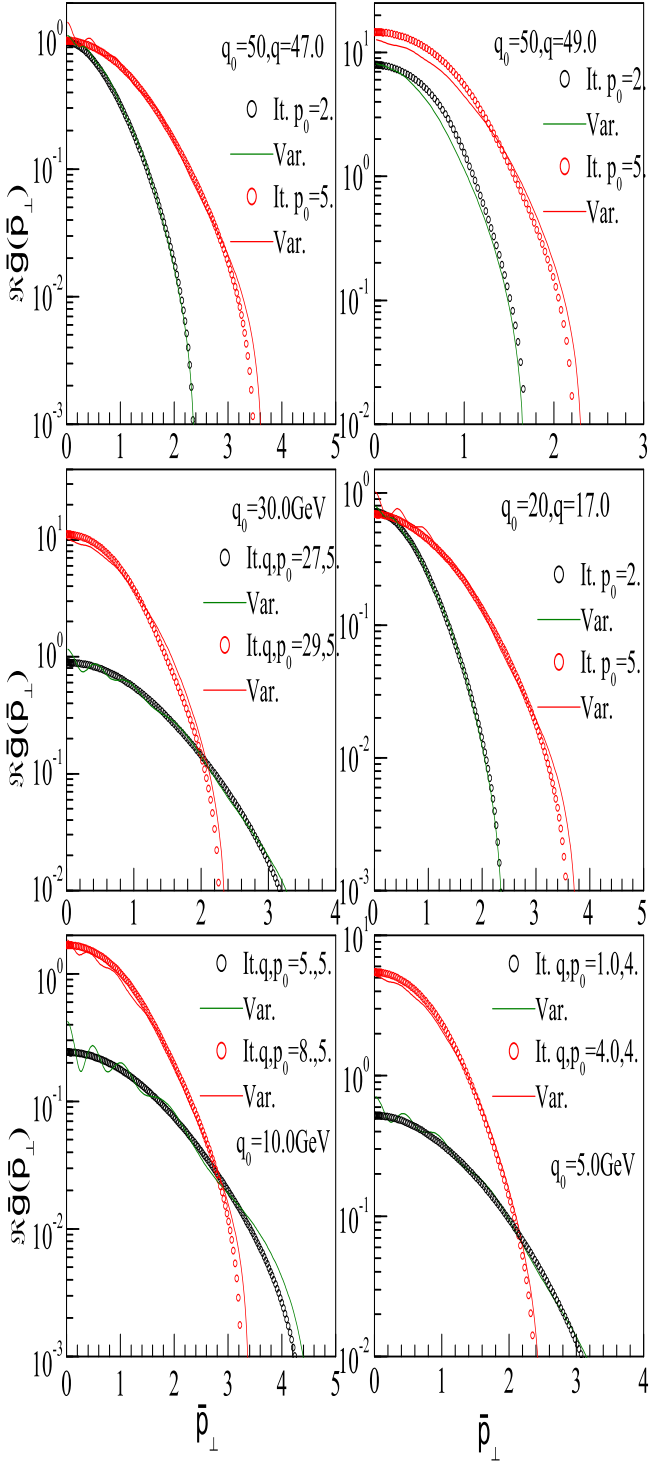}
\caption{   The  ${\bf p_\perp}$   distributions  $\Re{ \tilde{g}(\tilde{\bf p}_\perp)}$ (longitudinal part) of bremsstrahlung photon 
emission. The other details are as in  Figure. \ref{virtbt}.}
\label{virtbl}
\end{figure}
\begin{figure}[!]
\hspace{-1.0cm}
\includegraphics[height=21.cm,width=9.50cm]{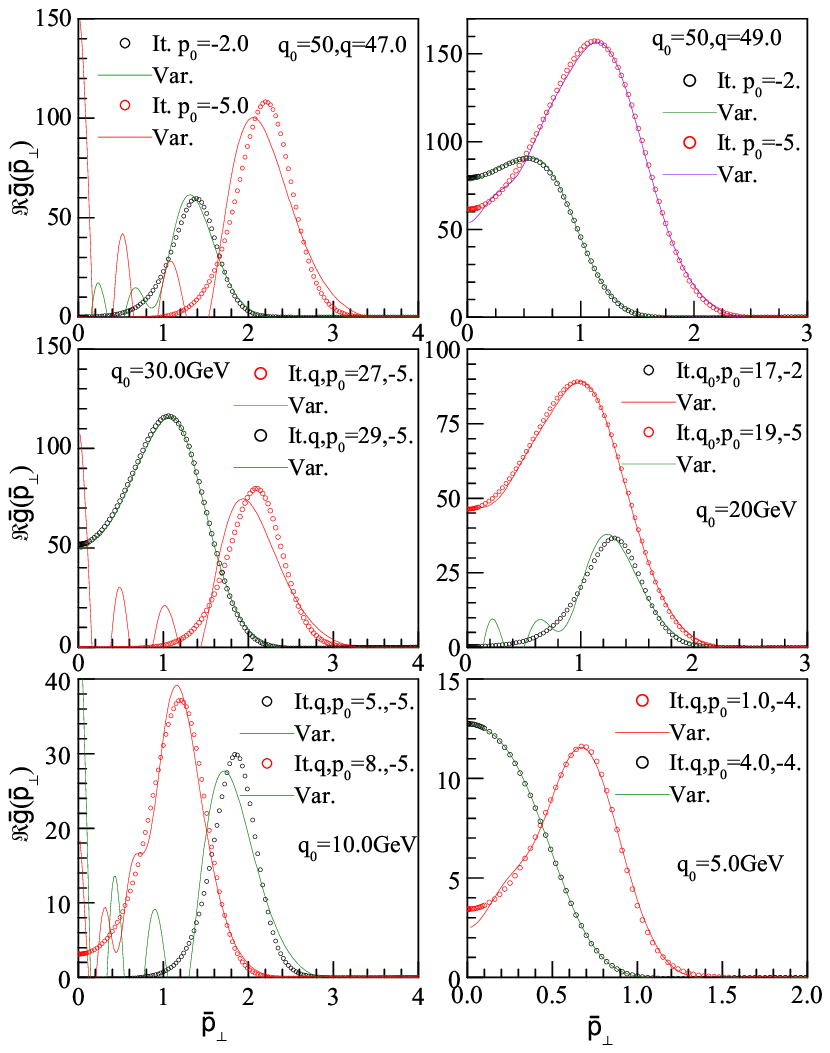}
\caption{  The ${\bf p_\perp}$   distributions $\Re{ \tilde{g}(\tilde{\bf p}_\perp)}$  (longitudinal part) of {\bf aws} process
The oter details are as in Figure. \ref{virtat}.}
\label{virtal}
\end{figure}
Figure \ref{virtat} shows the  ${\bf p_\perp}$   distributions  ${\bf \tilde{p}_\perp} \cdot\Re{\bf \tilde{f}(\tilde{p}_\perp)}$  of   ${\bf aws}$
  process for five values of $q_0=50,30,20,10,5$ GeV. For each $q_0$, similar to previous figure, distributions for two quark momenta values are shown .  
Figure \ref{virtbl} shows the  ${\bf p_\perp}$   distributions  ${\Re{\tilde{g}}({\bf \tilde{p}}_\perp)}$  of    
bremsstrahlung  process for five values of $q_0$. Figure \ref{virtal} shows the  ${\bf p_\perp}$   distributions for ${ \Re{\tilde{g}}({\bf \tilde{p}}_\perp)}$   
of (longitudinal  part of)  ${\bf aws}$  process for five values of $q_0$. \\
The  ${\bf p_\perp}$   distributions presented in Figures \ref{virtbt}-\ref{virtal} proved the validity of variational method. For low $Q^2$ this method  can 
be applied to study  the LPM effects in virtual photon emission.  However, the variational parameter needs to be optimized as it is not known for virtual photon 
case. For the case of real photons, we have shown an empirical estimate for  the optimized variational parameter in terms of real photon dynamical
variable $x_1$.  For the case of virtual photon emission such a simple empirical formula in terms of $x_1$ is not valid and  this should be a function also of 
$Q^2$. Therefore, in order to predict the optimized variational parameter for virtual photon emission case,  we propose to examine the  peak positions 
values  of $\bf p_\perp$ distributions from  iterations method. In Eqs.\ref{x0}-\ref{xbt} we define four dimensionless variables used in the following work.  
Especially,  $x^b_T$ of Eq.\ref{xbt} is the relevant dynamical variable for virtual photon emission at high $Q^2$ and the  variable  $x_1$  is for real photons.
(inverse  of  $x$  variable  used  in \cite{svsprc}).  We searched for the peak position values of ${\bf p_\perp}$ distributions  from the iterations method. 
These are functions of $p_0,q_0,Q^2$ and we searched for dynamical variables that  could represent the peak positions for virtual photon case. 
\begin{figure}[!]
\hspace{-0.80cm}
\includegraphics[height=16.cm,width=9.250cm]{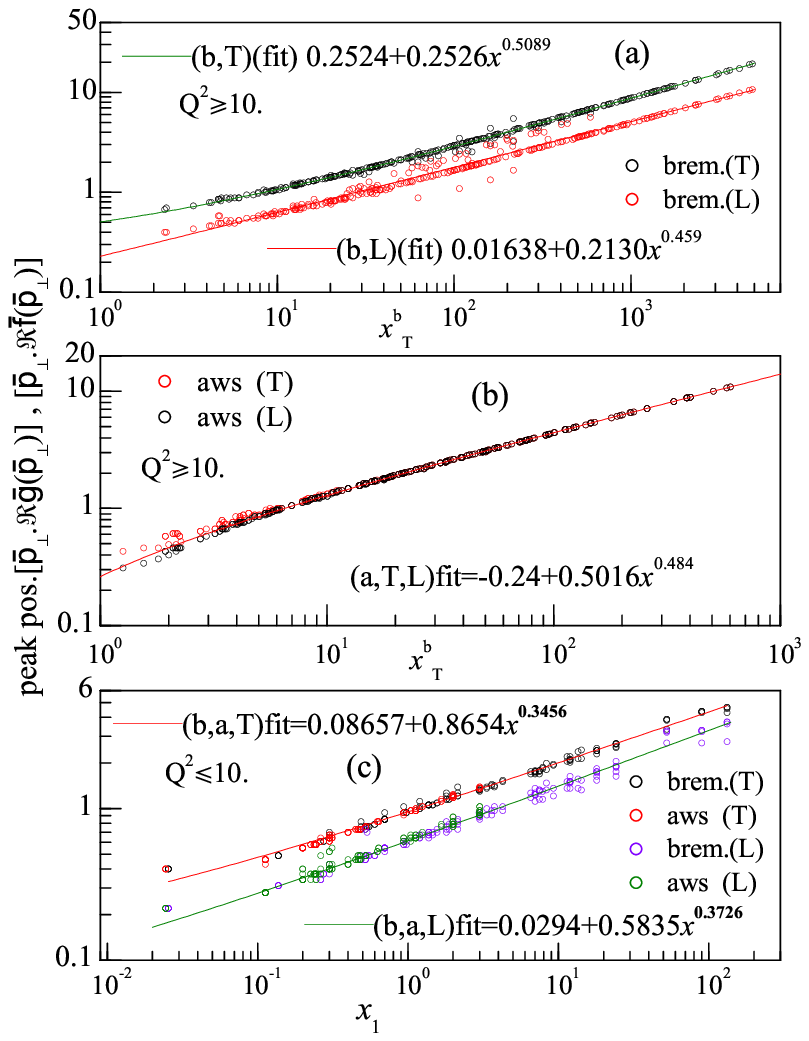}
\caption{ The peak positions of the   ${\bf  \tilde{p}_\perp}\cdot{\bf  \Re\tilde{f}}({\bf  \tilde{p}_\perp})$ and  
${ \tilde{p}_\perp}\cdot \Re\tilde{g}({\bf \tilde{p}_\perp})$ distributions  corresponding to various processes  and photon virtuality $Q^2$ 
and quark momenta.  All  data points shown are peak position results obtained from iterations method. The solid curves are the empirical  
power law fits shown on the figures. One should note that  the x-scale is $x^b_T$ for figures (a,b) and  scale is $x_1$ for figure (c)  (for details see text). 
These empirical fits give a rough estimate of the optimized variational parameter for use in variational method. The labels b,a represent the processes.
The lables T,L represent components.}
\label{virtppba}
\end{figure}
\begin{eqnarray}
x_0&=&\frac{|(p_0+q_0)p_0|}{q_0T  }  \label{x0} \\
x_1& = & x_0\frac{M_\infty^2}{m_D^2}  \label{x1} \\
x_2 &=& x_0\frac{Q^2}{q_0T}  \label{x2} \\
x^b_T& =& x_1+x_2     \label{xbt}   
\end{eqnarray}
Figure \ref{virtppba}(a) shows the peak positions of the  (transverse) ${\bf  \tilde{p}_\perp}\cdot{\bf  \Re\tilde{f}}({\bf  \tilde{p}_\perp})$  and  
(longitudinal) ${\tilde{p}_\perp}\Re\tilde{g}({\bf \tilde{p}_\perp})$ distributions for bremsstrahlung process at $Q^2\ge 10GeV^2$.  The data is generated  by 
solving the integral equations using iterations method for the set of values $\{p_0,q_0,Q^2\}$.  For the longitudinal part, the peak positions do not exist for 
several  values of $p_0,q_0,Q^2$ and  the corresponding   $\Re\tilde{g}({\bf \tilde{p}_\perp})$ distributions exhibit a Woods-Saxon form as can be seen in 
Figs.\ref{virtbl},\ref{virtal}. However, the  integrand of  $\int d^2{\bf \tilde{p}_\perp} \Re\tilde{g}({\bf \tilde{p}_\perp}),  ~{\it i.e.,}~
{\tilde{p}_\perp}\Re\tilde{g}({\bf \tilde{p}_\perp})$, exhibits peaking behaviour.  Therefore,  we propose that the variational paramter may be fixed as the peak positions of 
these respective distributions.  One should notice the $x$-scale  is $x^b_T$  as defined in Eq.\ref{xbt} for both transverse and longitudinal  distributions. The data 
is fitted by a  function whose formula and the  coefficients are mentioned in the figure (a) for both transverse and longitudinal parts.  We have chosen a fitting 
 function of type $y=a+bx^p$ because, as $x\rightarrow 0$, we want the peak positions to saturate as in real photon case (for real photons  $A_v\ge 0.31$).\\
 Figure \ref{virtppba}(b) shows the similar results for ${\bf aws}$ process   at $Q^2\ge 10GeV^2$.  Here again the $x$ scale is  $x^b_T$. Moreover, 
the  distributions  ${\bf  \tilde{p}_\perp}\cdot{\bf  \Re\tilde{f}}({\bf  \tilde{p}_\perp})$,  ${\tilde{p}_\perp}\Re\tilde{g}({\bf \tilde{p}_\perp})$ 
 have approximately same peak position values for different $p_0,q_0,Q^2$. The fit 
functions and coefficients are shown in Figure \ref{virtppba}(b).\\
Figure \ref{virtppba}(c) shows the peak positions   for bremsstrahlung and ${\bf aws}$ process  for  $Q^2\le 10GeV^2$.  In this
 figure, the transverse distributions ({\it i.e}.,  ${\bf  \tilde{p}_\perp}\cdot{\bf  \Re\tilde{f}}({\bf  \tilde{p}_\perp}))$  for bremsstrahlung and ${\bf aws}$ have 
approximately same peak positions  for  different $p_0,q_0,Q^2$ values. Notice that the $x$ scale is the $x_1$ variable defined in Eq.\ref{x1}. Surprisingly, 
all these peak position  values scale with  $x_1$ variable rather than the usual $x^b_T$  for transverse components.  This might be indicating that for small virtuality, 
the  relevant scale is $x_1$, coinciding with real photon scale. Similarly, the longitudinal components    ${\tilde{p}_\perp}\Re\tilde{g}({\bf \tilde{p}_\perp})$ have 
same peak  positions  for these two processes.  The fit functions and coefficients are shown in  figure. Using the formulae  given in  
Figures.\ref{virtppba}(a,b,c)  for the case of virtual photon  emission, one may choose the variational parameter   $A_v(p_0,q_0,Q^2)$ 
to be around these peak position values. \\
It is interesting to compare the results for real and virtual photon peak positions. The peak positions of 
${\bf  \tilde{p}_\perp}\cdot{\bf  \Re\tilde{f}}({\bf  \tilde{p}_\perp})$ distributions for real photons  are given by empirical fits in Fig.\ref{realppxtallba}(c).
The  peak positions of  ${\bf  \tilde{p}_\perp}\cdot{\bf  \Re\tilde{f}}({\bf  \tilde{p}_\perp})$, ~${ \tilde{p}_\perp} \Re\tilde{g}({\bf \tilde{p}_\perp})$ distributions for virtual photons
for $Q^2\le10GeV^2$ are given in Fig.\ref{virtppba}(c).  In Figure \ref{realvirt1}, we compare these three results. As seen in figure, the
peak positions for real photon and the transverse part of virtual photons are approximately the same. It should be noted that the numerical calculations have errors
such as the iterations method has convergence errors, peak search has errors, errors in empirical fit of peak positions etc. Further, $\alpha_s$ value used is 0.2 for real photons 
and 0.3 for virtual photon studies.  This close matching of peak positions of real and virtulal cases is  rather surprising, as a strong $Q^2$
dependence is expected for virtual photons.  Similar  surprising results were already presented in \cite{gef} regarding photon emission function.
\begin{eqnarray}
\Im{\Pi^\mu}_{R\mu} &\sim& \int_{-\infty}^\infty  dp_0 [n_F(r_0)-n_F(p_0)] \otimes \nonumber \\
&& \int \frac{d^2{\bf \tilde{p}_\perp}}{(2\pi)^2}\left[ \frac{p_0^2+r_0^2}{2(p_0r_0)^2} \Re{\bf \tilde{p}_\perp.\tilde{f}(\tilde{p}_\perp)} + \right. \nonumber \\
&& \left. \frac{1}{\sqrt{\left|p_0r_0\right|}}\frac{Q^2}{q^2} \left(\frac{1}{m_D}\right)\Re \tilde{g}({\bf\tilde{ p}_\perp})\right] 
\label{impolar}
\end{eqnarray}
\begin{figure}[!]
\hspace{-1.0cm}
\includegraphics[height=10.cm,width=9.50cm]{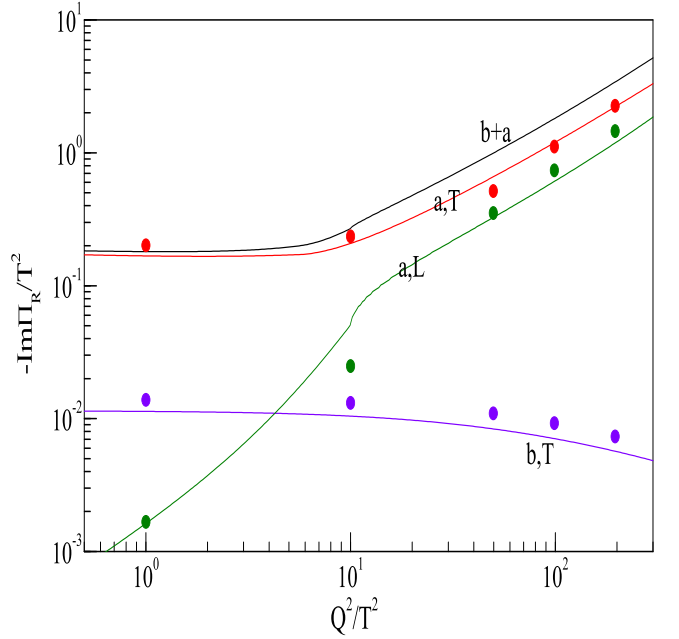}
\caption{$\Im \Pi_R$ plotted as a function of $Q^2/T^2$  for a photon energy of 50GeV. The transverse components of bremsstrahlung, $\bf aws$,  
the insignificant contribution from longitudinal parts are  shown. Symbols represent variational calculations. For  reference  the curves shown are 
taken from \cite{gef}.}
\label{imself50md1}
\end{figure}
In the previous section, we compared the results from variational  and  iterations methods and gave empirical fits to optimized values of variational parameters.  Using the 
variational method and the variational parameters given in Figures.\ref{virtppba}{a,b,c}, we repeated the variational calculations for all the values of $\{q_0,q,p_0\}$. 
These calculations cover  transverse and longitudinal parts of bremsstrahlung and $\bf aws$ processes for which we have reference distributions  from iterations method.
Now, we compared these two sets of $p_\perp$ distributions. It was noticed that the transverse bremsstrahlung distributions were exactly reproduced for all
$\{q_0,q,p_0\}$ values. The longitudinal contributions to bremsstrahlung were not not well reproduced, as the variational data showed oscillations around the iterations 
distributions. The transverse and longitudinal parts of $\bf aws$ were reasonably well reproduced for all $Q^2\le100GeV^2$.  The $\bf aws$ distributions showed
sensitivity to variational parameter.  Therefore, it should be noted that the empirical optimized variational parameter is a good approximation, however, 
one needs to vary the variational parameter around these empirical values to converge the variational results. \\
The imaginary part of retarded photon polarization tensor (represented by $\Im \Pi_R$)   is calculated using the $\bf p_\perp$ integrated values as 
given in Eq.\ref{impolar} (see Eq.16 of \cite{lpmdilep}).  The  required  $\bf p_\perp$ integrated values were generated using variational method
with empirical variational parameters.  In this Eq.\ref{impolar}, one should note   the factor $1/m_D$   in the longitudinal part for reasons explained before.  
All terms in this equation  contributing  to $\Im \Pi_R$ are calculated.
Figure \ref{imself50md1} shows the  $\Im \Pi_R$ plotted as a function of $Q^2/T^2$  for a photon energy of 50GeV.  The results of variational method are 
represented by symbols in the figure. The transverse components of bremsstrahlung and $\bf aws$ represented in figure by b,T and  a,T. The longitudinal  
contribution to   $\bf aws$ is shown as  a,L.  The results from variational method  (symbols) for various components have been normalized to approximately match 
with respective curves at $Q^2/T^2=1$.   At this photon energy (50GeV), the contribution of bremsstrahlung  is completely negligible. 
The bremsstrahlung is insignificant because the $q_0$ is very high. The transverse component of $\bf aws$ only contributes, with a small contribution from 
longitudinal component above $Q^2>20GeV^2$. The curves represent the imaginary polarization tensor by empirical method proposed in \cite{gef}. It can be 
seen that the variational has predicted reasonably  well the the results of \cite{gef}  for $Q^2/T^2$~in the range of ~$1-250$.  As shown in \cite{lpmdilep},  
the multiple re-scatterings  in the medium only marginally increase the  $\Im \Pi_R$ at low $Q^2$.  However, the re-scattering corrections  smooth out the 
discontinuity at the tree level threshold $Q^2=4M_\infty^2$ \cite{lpmdilep}.
\section{Conclusion}
\noindent
The  photon emission  processes from the quark gluon plasma have been studied as a function of photon mass, considering LPM suppression effects at  a  fixed
temperature of the plasma.  Self-consistent  iterations method and the variational method have been used to solve the AMY and AGMZ integral  equations.  
We  obtained the  ${\bf  \Re\tilde{f}}({\bf  \tilde{p}_\perp})$,  ${  \Re\tilde{g}}({\bf \tilde{p}_\perp})$ distributions as  a  function  of  photon  mass,  photon 
energy and  quark momentum.  The corresponding  distributions from  variational method have been compared with the results of iterations method  for validating the variational 
approach.  In order to fix variational parameter, the peak positions of the $\bf p_\perp$   distributions  have been studied in detail for both real and virtual photons 
emission using iteration method.  We identified relevant dynamical  scales for the peak position values for   real and virtual photon  cases.  The peak  positions have 
been represented  using appropriate dynamical scales and fitted  with empirical formulae.  In terms of these formulae, the optimized variational parameter  can be 
approximately estimated.  Using this empirical variational parameter, imaginary  part of  retarded photon polarization tensor has been calculated at photon energy of 50GeV.
\acknowledgements
I am thankful for discussions with Drs. A. K. Mohanty,  R. K. Choudhury, S. Kailas and  S. Ganesan. Computer Division of  BARC  is thanked for 
computational services provided. I  gratefully acknowledge  the co-operation extended to me by my wife  S.V. Ramalakshmi  during this study.
\noindent

\end{document}